\documentclass[11pt]{article}
\usepackage{amsfonts}
\usepackage{adjustbox}
\usepackage{mathrsfs}
\usepackage[latin1]{inputenc}
\usepackage{float}
\usepackage{amsmath,amssymb}
\usepackage{graphicx}
\usepackage{caption}
\usepackage{subcaption}
\usepackage{amsfonts}
\usepackage{latexsym}
\usepackage{epstopdf}
\usepackage{geometry} 
\usepackage{bbm} 
\usepackage[active]{srcltx}
\usepackage{graphicx}
\usepackage{epstopdf}
\usepackage{booktabs}
\usepackage{multirow}
\usepackage{siunitx}
\usepackage{enumitem}
\usepackage{scrextend}
\usepackage{lipsum}
\usepackage{setspace}
\usepackage[
bookmarks=true,         
bookmarksnumbered=true, 
colorlinks=true, pdfstartview=FitV, linkcolor=blue, citecolor=blue,
urlcolor=blue]{hyperref}

\providecommand{\U}[1]{\protect\rule{.1in}{.1in}}
\newtheorem {theorem}{Theorem}[section]

\newtheorem{definition}{Definition}[section]

\newcommand{\E}{\mathbb{E}}

\newcommand{\bi}[1]{\mbox{\boldmath{$ #1 $}}}

\begin{document}
\title{\Large Jackknife empirical likelihood confidence intervals for the categorical Gini correlation}
\vspace{0.5cm}
\author{Sameera Hewage and Yongli Sang\thanks{CONTACT: Yongli Sang; Email: yongli.sang@louisiana.edu}}
\date{%
Department of Mathematics, University of Louisiana at Lafayette, Lafayette, LA 70504, USA\\
\today
}

\maketitle

\begin{abstract}
The categorical Gini correlation, $\rho_g$, was proposed by Dang $et$ $al.$ \cite{Dang2021} to measure the dependence between a categorical variable, $Y$, and a numerical variable, $X$. It has been shown that $\rho_g$ has more appealing properties than existing dependence measurements. 
In this paper, we develop the jackknife empirical likelihood (JEL) method for $\rho_g$. Confidence intervals for the Gini correlation are constructed without estimating the asymptotic variance. Adjusted and weighted JEL are explored to improve the performance of the standard JEL. 
Simulation studies show that our methods are competitive to existing methods in terms of coverage accuracy and shortness of confidence intervals. The proposed methods are illustrated in an
application on two real datasets.

\end{abstract}
\noindent {\bf Keywords:} Categorical Gini correlation; Jackknife empirical likelihood; Wilk's theorem.
\noindent 

\vskip.2cm 
\noindent {\textit{MSC 2020 subject classification}: 62H12, 62H20}

\section{Introduction}
Categorical Gini correlation proposed by Dang $et$ $al.$ \cite{Dang2021} is a dependence measure between a numerical variable $\bi X$ and a categorical variable $Y$. 
Suppose that $\bi X$ is a numerical random variable from the distribution $F$ in $\mathbb{R}^d$. $Y$ is the categorical response variable taking values $L_1, ...,L_K$ and its distribution $P_Y$ is $P(Y = L_k) = p_k>0$ for $k=1,2,...,K$.
Assume that the conditional distribution of $\bi X$ given $Y=L_k$ is $F_k$. 
When the conditional distribution of $\bi X$ given $Y$ is the same as the marginal distribution of $\bi X$, $\bi X$ and $Y$ are independent. Otherwise, they are dependent. The categorical Gini covariance and correlation measure dependence based on the weighted distance between marginal and conditional distributions. 
Denote $\psi_k$ and $\psi$ as the characteristic functions of $F_k$ and $F$, respectively and define a weighted $L_2$ distance between $\psi_k$ and $\psi$ as
\begin{align} \label{charac}
&T(F_k, F)= c(d) \int_{\mathbb{R}^d} \frac{|\psi_k(\bi t) -\psi(\bi t)|^2}{\|\bi t\|^{d+1}} d\bi t,
\end{align}
where $c(d) = \Gamma((d+1)/2)/\pi^{(d+1)/2}$. Then, the Gini covariance between $\bi X$ and $Y$ is defined as 
\begin{equation}\label{Gcov}
\mbox{gCov}(\bi X,Y) = \sum_{k=1}^K p_k T(F_k, F).
\end{equation}
The Gini covariance measures dependence of $\bi X$ and $Y$ by quantifying the difference between the conditional and the unconditional characteristic functions. The corresponding Gini correlation standardizes the Gini covariance to have 
a range in [0,1]. 
%
%
When $d=1$, the categorical Gini covariance and correlation between $X$ and $Y$ can be defined by
\begin{align*}
&\mbox{gCov}( X,Y) = \sum_{k=1}^K p_k \int_{\mathbb{R}} \big(F_{k}(x)-F(x)\big)^2 dx, \\ 
&\rho_g( X,Y) =\dfrac{ \sum_{k=1}^K p_k \int_{\mathbb{R}} \big(F_{k}(x)-F(x)\big)^2 dx}{\int_{\mathbb{R}} F(x)(1-F(x)) \; dx},
\end{align*}
where the covariance is the weighted squared distance between the marginal distribution and the conditional distribution. It has been shown that $\rho_g$ (1) has a lower computational cost; (2) is more straightforward to perform statistical inference; (3) more robust to deal with unbalanced data than the popular distance correlation \cite{Szekely07}. These appealing properties motivate us to develop inference of the categorical Gini correlation.

Dang $et$ $al.$ \cite{Dang2021} estimated the categorical Gini covariance and correlation by $V$-statistics and established asymptotic distributions of the $V$-estimators. They admit normal limits when $\bi X$ and $Y$ are dependent.
However, the asymptotic variance for the estimator is difficult to compute. In this paper, we develop a nonparametric method to build confidence intervals for the categorical Gini correlation without estimating the asymptotic variance. 

In fact, $T(F_k, F)$ in (\ref{charac}) is the energy distance defined in Sz\'{e}kely \& Rizzo \cite{Szekely2013a, Szekely17} that can be written as 
\begin{align*}
T(F_k, F)= 2 \E\|\bi X_1^{(k)} -\bi X_1\|-\E\|\bi X_1^{(k)} -\bi X_2^{(k)}\| -\E \|\bi X_1-\bi X_2\|,
\end{align*}
where $(\bi X_1,\bi X_2)^T$ and $(\bi X_1^{(k)}, \bi X_2^{(k)})^T$ are independent pair variables independently from $F$ and $F_k$ respectively. Therefore, the Gini covariance defined by (\ref{Gcov}) is a weighted average of energy distance between $\bi X^{(k)}$ and $\bi X$. 
And it can be naturally estimated by a function of $U$-statistics. 

Jackknife empirical likelihood (JEL) is a nonparametric method proposed by Jing, Yuan and Zhou \cite{Jing2009} to overcome the computational burden of empirical likelihood (EL) \cite{Owen1988, Owen1990} when $U$-statistics are involved. It combines the jackknife and EL by applying EL to the jackknife pseudo-values. JEL has been applied in numerous problems since its introduction. It has been applied to constructing confidence intervals for ROC curve \cite{Gong2010, Yang2013, Yang2015}, Gini index \cite{Wang2016}, Gini correlations \cite{Sang2019}, Spearman's rho \cite{Wang2011}, quantile \cite{Yang2017, Yang2018} and testing two-sample \cite{Feng2012, Chen2015} and K-sample problems \cite{Sang2021a, Sang2021b}. One can refer to Liu and Zhao \cite{Liu2022} for more applications of JEL. In this paper, we apply JEL  to build confidence intervals for the categorical Gini correlation. Chen
et al. \cite{Chen2008} added two artificial points to the original pseudo-value data set and developed the balanced augmented
JEL to improve the performance of JEL. To solve the sensitivity problems with outliers, Sang et al. \cite{Sang2020} derived the weighted JEL by assigning smaller weights to outliers, thus making JEL more robust. We also explore the adjusted JEL and the robust JEL for the categorical Gini correlation. 


The remainder of the paper is organized as follows. In Section \ref{sec:jel}, we develop the JEL for the categorical Gini correlation.
In Section \ref{sec:simulationstudy}, we conduct simulation studies to evaluate the performance of the JEL methods. Real data analysis for two datasets is illustrated in Section \ref{sec:realdata} to compare the proposed procedure with currently available approaches. 
We conclude and discuss future works in Section \ref{sec:conclusion}. Some detailed derivations of Remarks and all technical proofs are provided in Appendix.

\section{JEL for the categorical Gini correlation}\label{sec:jel}

\subsection{Categorical Gini correlation}
The Gini covariance and correlation can be represented in terms of the multivariate Gini mean differences (GMD). 
The GMD is an alternative measure of variability. Let $(\bi X_1,\bi X_2)^T$ and $(\bi X_1^{(k)}, \bi X_2^{(k)})^T$ be independent pair variables independently from $F$ and $F_k$, respectively.
The GMDs for $F$ and $F_k$ are defined by 
\begin{align*}
\Delta =\E\|\bi X_1-\bi X_2\|, \quad \Delta_k=\E \|\bi X_1^{(k)}-\bi X_2^{(k)}\|.
\end{align*}
From \cite{Dang2021}, we have
\begin{equation*}\label{gcov}
\mbox{gCov}(\bi X,Y) = \Delta-\sum_{k=1}^Kp_k\Delta_k,
\end{equation*}
and 
\begin{equation} \label{mgc}
\rho_g(\bi X, Y) = \frac{ \Delta-\sum_{k=1}^Kp_k\Delta_k}{\Delta}. 
\end{equation}
We can see that the Gini correlation is the ratio of between variation and overall variation. 



Dang {\em et al.} \cite{Dang2021} used V-statistic estimators and derived limiting distributions of the estimators under the classical setting when the dimension of $\bi X$ is fixed. More specifically, 
suppose a sample ${\cal D} =\{(\bi X_1, Y_1), (\bi X_2, Y_2), ...., (\bi X_n, Y_n)\}$ is drawn from the joint distribution of $\bi X$ and $Y$ . We can write ${\cal D} ={\cal D}_1\cup {\cal D}_2...\cup {\cal D}_K$, where ${\cal D}_k=\left \{\bi X^{(k)}_{1}, \bi X^{(k)}_{2}, ...,\bi X^{(k)}_{n_k}\right \}$ is the sample with $Y_i=L_k$ and $n_k$ is the number of sample points in the $k^{th}$ class. 
Dang $et$ $al$. \cite{Dang2021} estimated the Gini correlation for (\ref{mgc}) as 
\begin{align*} 
\hat{\rho}_g(\bi X,Y) = 1-\frac{\sum_{k=1}^K \hat p_k \tilde{\Delta}_k}{\tilde{\Delta}}=\frac{\tilde{\Delta}- \sum_{k=1}^K \hat p_k \tilde{\Delta}_k}{\tilde{\Delta}}, 
\end{align*}
where $\hat{p}_k=\dfrac{n_k}{n},$ and 
\begin{align} \label{Vstat}
& \;\; \tilde{\Delta}_k = n_k^{-2}\sum_{1 \leq i,j \leq n_k} \|\bi X_i^{(k)} -\bi X_j^{(k)}\|, \;\;\; \tilde{\Delta} = n^{-2} \sum_{1\leq i,j \leq n} \|\bi X_i -\bi X_j\|.
\end{align}
 The estimators in (\ref{Vstat}) are $V$ Statistics, which are biased. They worked with
biased sample versions to avoid dealing with complicated constants in the ensuing result of $\hat{\rho}_g(\bi X,Y)$. They have shown that if $\mbox{gCor}(\bi X,Y)\neq 0$, then 
\begin{equation} \label{eqn:norm}
\sqrt{n} (\hat{\rho}_g(\bi X,Y) -\mbox{gCor}(\bi X,Y)) \stackrel{D}{\longrightarrow} {\cal N}(0, \sigma_g^2),
\end{equation}
where $\sigma_g^2$ is the asymptotic variance. 

Confidence intervals for $\rho_g$ can be constructed based on the asymptotic normality. However, the variance, $\sigma^2_g$, has complicated form and is difficult to compute. 
An estimate of $\sigma^2_g$ is needed either by a Monte Carlo simulation or based on the jackknife method. 
Let $\hat {\rho_g}_{(-i)}$ be the jackknife pseudo value of the Gini correlation estimator $\hat {\rho_g}$ based on the sample with the $i^{th}$ observation deleted. Then, the jackknife estimator of $\sigma^2_g$ is
\begin{align}\label{jel_var}
\widehat{\sigma}_g^2 =\dfrac{n-1}{n}\sum_{i=1}^n (\hat{\rho}_{g(-i)} -\bar{\hat{\rho}}_{g(\cdot)})^2,
\end{align}
where $\bar{\hat{\rho}}_{g(\cdot)} =\dfrac{1}{n} \sum_{i=1}^n {\hat{\rho}}_{g(-i)},$ see \cite{Shao96}.

We will develop a nonparametric procedure, JEL, to build confidence intervals without estimating the complicated variance.


\subsection{JEL for the Categorical Gini correlation}\label{sec:cgc}
In order to apply JEL, we will use the $U$-estimators.
Define $h(x, y)=\|x-y\|$ and let \begin{align*}
&U_n={n \choose 2}^{-1}\sum_{1\leq i<j \leq n}h(\bi X_i, \bi X_j),\\
&U_{n_k}={n_k \choose 2}^{-1}\sum_{1\leq i<j \leq n_k}h(\bi X^{(k)}_i, \bi X^{(k)}_j),
\end{align*}
which could estimate the GMDs unbiasedly. That is,
\begin{align*}
\E U_n=\Delta, \quad \E U_{n_k}=\Delta_k, \; k=1,..,K. 
\end{align*}
Thus, the categorical Gini correlation (\ref{mgc}) can be written as 
\begin{align*}
\rho_g(\bi X, Y) = \frac{\E U_n-\sum_{k=1}^Kp_k\E U_{n_k}}{\E U_n}. 
\end{align*}

Define
\begin{align}\label{wn}
W_n(\rho_g)=U_n(1-\rho_g)-\sum_{k=1}^K \hat{p}_kU_{n_k}.
\end{align}
It is easy to have $\mathbb{E}W_n(\rho_g)=0$.

To apply the JEL to $W_n(\rho_g)$, we define the jackknife pseudo sample as 
\begin{align}\label{wn_pseudo}
\hat{V}_i(\rho_g)=nW_n(\rho_g)-(n-1)W^{(-i)}_{n-1}(\rho_g),
\end{align}
where $W^{(-i)}_{n-1}(\rho_g)$ is based on the sample with the $i^{\text{th}}$ observation $X_i$ being deleted. We have shown that $W_n(\rho_g)=\dfrac{1}{n}\sum_{i=1}^n\hat{V}_i(\rho_g)$ in the Appendix.

Let $\bi{\pi}=(\pi_1,...,\pi_n)$ be nonnegative numbers such that $\sum_{i=1}^{n}\pi_i=1.$ Then following the standard empirical likelihood method for a univariate mean over the jackknife pseudo-values (\cite{Owen1988}, \cite{Owen1990}), we define the JEL ratio at $\rho_g$ as
\begin{align*} 
R(\rho_g)=\max \left \{\prod_{i=1}^{n}(n\pi_i): \pi_i \ge 0, i=1,...,n; \sum_{i=1}^n \pi_i=1; \sum_{i=1}^n \pi_i \hat{V}_{i}(\rho_g)=0 \right \}.
\end{align*}
Utilizing the standard Lagrange multiplier technique, the jackknife empirical log-likelihood ratio at $\rho_g$ is
\begin{align*}
\log R(\rho_g)=-\sum_{i=1}^{n}\log[1+\lambda \hat{V}_{i}(\rho_g) ],
\end{align*}
where $\lambda=\lambda(\rho_g)$ satisfies 
\begin{align}\label{lambda}
\frac{1}{n}\sum_{i=1}^n \frac{ \hat{V}_{i}(\rho_g)}{1+\lambda \hat{V}_{i}(\rho_g)}=0.
\end{align}

Assume 
\begin{description} \label{Condition}
\item \textbf{C}1. $\E \|\bi X\|^2 < \infty$;
\item \textbf{C}2. $\rho_g(\bi X, Y) \neq 0$;
\item \textbf{C}3. $\dfrac{n_k}{n} \to \alpha_k, k=1,...,K$ and $\alpha_1+\alpha_2+...+\alpha_K=1$.
\end{description}
Note that condition \textbf{C}3 shows that none of the class from the $K$ groups can dominate the others.  We have the following Wilks' theorem.
\begin{theorem}\label{wilkrho}
Under conditions \textbf{C}1-\textbf{C}3, we have 
$$ -2 \log R(\rho_g)\stackrel{d}{\rightarrow} \chi^{2}_1, \;\;\;\text{as $n \to \infty$}.$$
\end{theorem}
Based on the theorem above, a $100(1-\alpha)\%$ jackknife empirical likelihood confidence interval for $\rho_g$ can be constructed as
\begin{align*}
I_{\alpha} = \{ \tilde{\rho_g}: -2 \log \hat{R}(\tilde{\rho_g})\le \chi^2_{1, 1-\alpha} \},
\end{align*}
where $\chi^2_{1, 1-\alpha}$ denotes the $100(1-\alpha)\%$ quantile of the chi-square distribution with one degree of freedom, and $ \log \hat{R}(\tilde{\rho_g})$ is the observed empirical log-likelihood ratio at $\tilde \rho_g$.



In application, an under-coverage problem may appear when the sample size is relatively small. In order to improve coverage probabilities, we utilize the adjusted empirical likelihood method \cite{Chen2008} by adding one more pseudo-value 
\begin{align}\label{pseudo}
\hat{V}_{n+1}(\rho_g)=-\frac{a_n}{n}\sum^{n}_{i=1}\hat{V}_{i}(\rho_g),
\end{align}
where $a_n=o_{p}(n^{2/3}).$

Sang, Dang and Zhao (\cite{Sang2020}) proposed weighted JEL to make the methodology more robust. 
\begin{definition}
Suppose that $\bi X_i$ ($i=1, ..., n$) are independently distributed from an unknown distribution $F$ with a $r$-dimensional parameter $\bi \theta$. Assume that $\pi_i$ is the probability mass placed on $\bi X_i$. Given a weight vector $\bi \omega_n$ with $\sum_{i=1}^{n}\omega_{ni}=1$ and $\omega_{ni} \geq0$, the weighted jackknife empirical likelihood (WJEL) for parameter $\bi \theta$ is then defined as

\begin{align*}
\text{\mbox{WJEL}}(\bi \theta)=\sup\left \{\prod_{i=1}^{n}\pi^{n\omega_{ni}}_i: \sum_{i=1}^{n} \pi_i=1, \sum_{i=1}^{n} \pi_i\hat{\bi V}_i(\bi \theta)=\bi 0 \right\},
\end{align*}
where $\hat{\bi V}_i(\bi \theta), i=1,...,n$ are the jackknife pseudo values. 
\end{definition}

With the same argument as JEL, the corresponding log-likelihood ratio admits a weighted Chi-squared distribution. 
Let $\bi \theta_0$ be the true value of $\bi \theta$ and $c=\lim_{n \to \infty}\sum_{i=1}^n n \omega^2_{ni}$. Under some mild regularity conditions stated in the Appendix, the Wilks theorem holds for the $U$-type WJEL ratio,

\begin{align*}
&\dfrac{l(\bi \theta_0)}{\sum_{i=1}^n n \omega^2_{ni}} \stackrel{d}{\rightarrow} \chi^2_r\;\; \text{as}\;\; n\to \infty,
\end{align*}

where 
\begin{align*}
l(\bi \theta) =-2 \log R(\bi \theta)=2\sum_{i=1}^{n} n\omega_{ni} \log \{1+\bi \lambda^T \hat{ \bi V}_i(\bi \theta) \}.
\end{align*}


Suggested by the authors in \cite{Sang2020}, the weight can be chosen by
\begin{align}\label{sw}
\omega_{ni}=\dfrac{D(\bi X_i; F_n)}{\sum_{j=1}^n D(\bi X_j; F_n)},
\end{align}
where $D(\bi x; F)= 1- \| \mathbb E_F \bi S(\bi x -\bi X) \|$ is the spatial depth function and 
$ D(\bi x, F_n) = 1- \left \|\frac{1}{n} \sum_{i=1}^n \frac{\bi x -\bi X_i}{\|\bi x -\bi X_i\|} \right\|$ is the sample counterpart. 



\section{Simulation study}\label{sec:simulationstudy}


In order to assess the proposed JEL confidence intervals, four groups of simulation studies are conducted to investigate the performance of:

\begin{labeling}[~-]{Second}
\item[\textbf{JEL}] the proposed JEL method. Confidence interval is constructed based on the asymptotic Chi-squared distribution;
\item[\textbf{AJEL}] Adjusted JEL by adding one more pseudo-value (\ref{pseudo}); Under the recommendation of \cite{Chen2008}, we take $a_n=\max (1, \log (n)/2)$;
\item[\textbf{WJEL}] Weighted JEL by using the spatial depth based weights (\ref{sw});
\item[\textbf{JV}] The confidence interval constructed based on the asymptotic normality in (\ref{eqn:norm}) with the asymptotic variance estimated by (\ref{jel_var}).
\end{labeling}

Random samples for the simulations are generated from several mixtures of normal distributions and exponential distributions. First two simulations are for the univariate cases ($d=1$) and the last two are for the multivariate cases ($d=5$).
We generate $3,000$ samples from each scenario and then we repeat each procedure $30$ times. 
%
%
%
%
%
%
%

For the first simulation, we consider $K=2, d=1$. The following scenarios with balanced $\bi p = (p_1, p_2) = (1/2, 1/2)$ and unbalanced $\bi p = (2/5, 3/5)$ of the total sample sizes of ($n = 60, n = 120$) are considered.
\begin{itemize}
\item $X \sim p_1\mathcal{N}(0, 1)+p_2\mathcal{N}(3, 1)$
\item $X \sim p_1\mathcal{N}(0, 1)+p_2\mathcal{N}(0, 3^2)$
\item $X \sim p_1exp(1)+p_2exp(4)$
\end{itemize}
The average coverage probabilities and average lengths as well as their standard deviations (in parenthesis) of 95\% confidence intervals are presented in Table \ref{tab:Table1}. In addition, we have provided the categorical Gini correlation value for each scenario by using the analytical formulas from examples in \cite{Dang2021}:
\begin{align*}
\rho_g( X,Y) =\dfrac{p(1 - p)(\theta - \beta)^2}{(2p - p^2)\theta^2 + (1 - p^2)\beta^2+ (1 - 2p + 2p^2)\theta \beta}
\end{align*}
if $F_1 = \exp(\theta)$ and $F_2 = \exp(\beta)$;
\begin{align*}
\rho_g( X,Y) =\dfrac{p(1 - p)[2a\Phi(a/\sqrt{2}) + 2\sqrt{2}\phi(a/\sqrt{2}) - a - 2/\sqrt{\pi}]}{(p^2 + (1 - p)^2)/\sqrt{\pi} + p(1 - p)[2a\Phi(a/\sqrt{2}) + 2\sqrt{2} \phi(a/\sqrt{2}) - a]}
\end{align*}
where $\phi(x)$ and $\Phi(x)$ are the density and cumulative functions of the standard normal distribution, respectively, if $F_1 = \mathcal{N}(\mu_1, \sigma^2), F_2 = \mathcal{N}(\mu_2, \sigma^2),$ and $a = |\mu_1 - \mu_2|/\sigma$;
\begin{align*}
\rho_g( X,Y) =\dfrac{p(1 - p)(\sqrt{2(1 + r^2)} - 1 - r)}{p^2 + (1 - p)^2r + p(1 - p)\sqrt{2(1 + r^2)}},
\end{align*}
if $F_1 = \mathcal{N}(\mu, \sigma_1^2), F_2 = \mathcal{N}(\mu, \sigma_2^2),$ and $r = \sigma_2/\sigma_1$. 
However, for the other cases in Table \ref{tab:Table2} - Table \ref{tab:Table4} where $K>2$ or $d>1$, it is extremely difficult to develop the analytical formula for the categorical Gini correlation. Thus, we utilized Monte Carlo simulation to approximate the population value with sample sizes that are in multiples of 1,000,000.


\begin{table}[H]
\caption{Coverage probabilities (standard deviations) and average lengths (standard deviations) of the categorical Gini correlations' interval estimators under different distributions when $K=2$ and $d=1$.}
\center
\scriptsize
\renewcommand{\arraystretch}{1.1}
\begin{tabular}{lllcccccccccr }
\hline \hline
Distribution & Parameter &{Method} & \multicolumn{2}{c}{$n=60$} & \multicolumn{2}{c} {$n=120$} \\

&&&CovProb & Length &CovProb & Length\\
\hline

& &JEL &.9418(.0031) &.2299(.0005) &.9462(.0042)&.1631(.0004)\\

& &AJEL &.9499(.0028) & .2328(.0005) &.9509(.0044)&.1644(.0004)\\

$\dfrac{1}{2}\mathcal{N}(0,1)+\dfrac{1}{2}\mathcal{N}(3, 1)$ 
& 0.4556 &WJEL &.9428(.0023) & .2155(.0005) &.9465(.0055)&.1511(.0004)\\

&&JV &.9342(.0060) & .1926(.0004) &.9431(.0044)&.1351(.0003)
\\ \cline{1-7}

&&JEL &.9428(.0023)&.1086(.0005) &.9476(.0038) &.0720(.0002) \\

&&AJEL &.9490(.0015)& .0865(.0002) & .9512(.0034)&.0591(.0001) \\

$\dfrac{1}{2}\mathcal{N}(0,1)+\dfrac{1}{2}\mathcal{N}(0, 3^2)$ 
&0.0557&WJEL &.9422(.0024) & .1064(.0007) &.9484(.0045)&.0664(.0004) \\

&&JV &.9558(.0027) & .0966(.0005) &.9556(.0035)&.0598(.0003) \\
\cline{1-7}

&&JEL &.9308(.0061) &.2430(.0006) &.9435(.0056)&.1699(.0004)\\

&&AJEL &.9378(.0054)& .2420(.0004) &.9473(.0056)&.1683(.0004)\\

$\dfrac{1}{2}exp(1)+\dfrac{1}{2}exp(4)$
&0.1525&WJEL &.9309(.0046)& .2268(.0006) &.9438(.0061)&.1566(.0005)\\

&&JV &.9438(.0040) & .2023(.0004) &.9498(.0052)&.1397(.0003)\\
\cline{1-7}

&&JEL &.9606(.0036) &.2626(.0005) &.9671(.0032)&.1850(.0002)\\

&&AJEL &.9670(.0032)& .2590(.0006) &.9701(.0031)&.1851(.0002)\\
$\dfrac{1}{3}\mathcal{N}(0,1)+\dfrac{2}{3}\mathcal{N}(3,1)$
& 0.4267 &WJEL &.9410(.0050)& .2591(.0006) &.9040(.0056)&.1842(.0003)\\

&&JV &.9590(.0033) & .2168(.0004) &.9645(.0029)&.1513(.0002)\\
\cline{1-7}

&&JEL &.9589(.0020) &.0860(.0004) &.9654(.0026)&.0563(.0001)\\
&&AJEL &.9637(.0024)& .0674(.0004) &.9681(.0023)&.0416(.0002)\\

$\dfrac{1}{3}\mathcal{N}(0,1)+\dfrac{2}{3}\mathcal{N}(0,3^2)$
&0.0430&WJEL &.9559(.0026)& .0818(.0006) &.9605(.0033)&.0505(.0003)\\

&&JV &.9674(.0019) & .0734(.0002) &.9706(.0027)&.0462(.0001)\\
\cline{1-7}

&&JEL &.9490(.0049) &.1992(.0007) &.9574(.0029)&.1380(.0003)\\

&&AJEL &.9547(.0046)& .1891(.0005) &.9607(.0030)&.1259(.0003)\\

$\dfrac{1}{3}exp(1)+\dfrac{2}{3}exp(4)$
&0.1176&WJEL &.9423(.0042)& .1930(.0005) &.9404(.0044)&.1344(.0005)\\

&&JV &.9571(.0038) & .1625(.0004) &.9619(.0024)&.1124(.0002)\\
\cline{1-7}

\hline\hline
\end{tabular}
\label{tab:Table1}
\end{table}

From Table \ref{tab:Table1}, we observe that for balanced cases with the normal mixtures, all methods keep good coverage probabilities, but the JEL methods produce better coverage probabilities when sample size is large ($n = 120$). For unbalanced cases with the normal mixtures, almost all methods suffer from a slight over-coverage problem thus resulting in conservative confidence intervals. The WJEL keeps well the coverage probability, while the JEL suffers from a slight over-coverage problem. In the balanced case, for exponential mixtures, we have good coverage probabilities and shorter confidence intervals for all methods when sample sizes increase. In the unbalanced case, for exponential mixtures, JEL methods perform better in terms of the coverage probability. As the sample size increases, the length of the confidence interval decreases for all methods.

In the second simulation study, we consider $K=3, d=1$. The following scenarios with balanced $\bi p = (p_1, p_2, p_3) = (1/3, 1/3, 1/3)$ and unbalanced $\bi p = (3/12, 4/12, 5/12)$ of the total sample sizes of ($n = 60, n = 120$) are considered. 
\begin{itemize}
\item $X \sim p_1\mathcal{N}(0, 1)+p_2\mathcal{N}(1, 1)+p_3\mathcal{N}(2, 1)$
\item $X \sim p_1\mathcal{N}(0, 1)+p_2\mathcal{N}(0, 2)+p_3\mathcal{N}(0, 3)$
\item $X \sim p_1exp(1)+p_2exp(2)+p_3exp(3)$
\end{itemize}
Table \ref{tab:Table2} presents the average coverage probabilities and average lengths as well as their standard deviations (in parenthesis) of 95\% confidence intervals.

\begin{table}[H]
\caption{Coverage probabilities (standard deviations) and average lengths (standard deviations) of the categorical Gini correlations' interval estimators under different distributions when $K=3$ and $d=1$.}
\center
\scriptsize
\renewcommand{\arraystretch}{1.1}
\begin{tabular}{lllcccccccccr }
\hline \hline
Distribution&Parameter&{Method}& \multicolumn{2}{c}{$n=60$}& \multicolumn{2}{c} {$n=120$} \\

&&&CovProb & Length &CovProb & Length\\
\hline

&&JEL &.9453(.0021) &.3066(.0005) &.9507(.0039)&.2124(.0002)\\

&&AJEL &.9527(.0030)& .3024(.0006) &.9560(.0043)&.2130(.0003)\\

$\dfrac{1}{3}\mathcal{N}(0, 1)+\dfrac{1}{3}\mathcal{N}( 1, 1)+\dfrac{1}{3}\mathcal{N}(2, 1)$
&0.2295&WJEL &.9472(.0031)& .2877(.0008) &.9498(.0047)&.1996(.0007)\\

&&JV &.9487(.0032) & .2536(.0004) &.9534(.0028)&.1739(.0002)\\
\cline{1-7}

&&JEL &.9458(.0041) &.3083(.0006) &.9522(.0030)&.2136(.0002)\\

&&AJEL &.9520(.0033)& .2935(.0015) &.9563(.0036)&.2097(.0005)\\
$\dfrac{3}{12}\mathcal{N}(0, 1)+\dfrac{4}{12}\mathcal{N}(1, 1)+\dfrac{5}{12}\mathcal{N}(2, 1)$
&0.2227&WJEL &.9469(.0040)& .2896(.0009) &.9489(.0026)&.2011(.0007)\\

&&JV &.9487(.0041) & .2544(.0005) &.9538(.0034)&.1745(.0002)\\
\cline{1-7}

&&JEL &.9506(.0024) &.1136(.0006) &.9537(.0033)&.0658(.0005)\\
&&AJEL &.9561(.0027)& .0929(.0004) &.9573(.0032)&.0519(.0003)\\

$\dfrac{1}{3}\mathcal{N}(0, 1)+\dfrac{1}{3}\mathcal{N}(0, 2)+\dfrac{1}{3}\mathcal{N}(0, 3)$
&0.0392&WJEL &.9544(.0033)& .1285(.0008) &.9568(.0028)&.0726(.0005)\\
&&JV &.9580(.0026) & .1189(.0005) &.9596(.0029)&.0665(.0004)\\
\cline{1-7}

&&JEL &.9573(.0036) &.1061(.0007) &.9608(.0033)&.0619(.0003)\\

&&AJEL &.9619(.0031)& .0851(.0005) &.9633(.0027)&.0474(.0003)\\

$\dfrac{3}{12}\mathcal{N}(0, 1)+\dfrac{4}{12}\mathcal{N}(0, 2)+\dfrac{5}{12}\mathcal{N}( 0, 3)$
& 0.0341&WJEL &.9608(.0029)& .1154(.0008) &.9641(.0023)&.0648(.0003)\\
&&JV &.9627(.0028) & .1063(.0006) &.9646(.0031)&.0601(.0003)\\
\cline{1-7}

&&JEL &.9180(.0048) &.2229(.0010) &.9306(.0053)&.1523(.0005)\\

&&AJEL &.9241(.0049)& .1945(.0011) &.9346(.0049)&.1342(.0007)\\

$\dfrac{1}{3}exp(1)+\dfrac{1}{3}exp(2)+\dfrac{1}{3}exp(3)$
&0.0784&WJEL &.9186(.0064)& .2094(.0009) &.9288(.0042)&.1393(.0004)\\
&&JV &.9287(.0044) & .1883(.0009) &.9392(.0057)&.1249(.0003)\\
\cline{1-7}

&&JEL &.9252(.0039) &.2033(.0010) &.9359(.0057)&.1382(.0004)\\

&&AJEL &.9316(.0033)&.1752(.0013)&.9396(.0052)& .1201(.0008) \\
$\dfrac{3}{12}exp(1)+\dfrac{4}{12}exp(2)+\dfrac{5}{12}exp(3)$
&0.0686&WJEL &.9260(.0041)& .1920(.0009) &.9313(.0033)&.1276(.0006)\\

&&JV &.9376(.0031) & .1708(.0008) &.9429(.0046)&.1129(.0003)\\
\cline{1-7}

\hline\hline
\end{tabular}
\label{tab:Table2}
\end{table}
We observe that under the normal mixtures, all methods have good coverage probabilities and the JEL methods produce better coverage probabilities when sample size is large ($n = 120$).
For the exponential mixtures, all the methods suffer from the under-coverage problem. However, as the sample size increases, the issue becomes less serious. AJEL and JV perform closely and have better  coverage probabilities. The confidence interval gets shorter for all methods as the sample size increases.

For the third simulation study, we consider  the mixture of following multivariate distributions  with balanced $\bi p = (p_1, p_2) = (1/2, 1/2)$ and unbalanced $\bi p = (2/5, 3/5)$ of the total sample sizes of ($n = 100, n = 200$). 
\begin{itemize}
\item $\bi X \sim p_1\mathcal{N}(\bi 0_d, \bi I_{d\times d})+p_2\mathcal{N}(\bi 1_d, 2 \bi I_{d\times d})$,
\item $\bi X \sim p_1\mathcal{N}(\bi 0_d, \bi I_{d\times d})+p_2\mathcal{N}(\bi 0_d, 3\bi I_{d\times d})$,
\item $\bi X \sim p_1\textrm{Exp}(\bi 1_d)+p_2\textrm{Exp}(\bi 3_d)$,
\end{itemize}
where $\bi 0_d$, $\bi 1_d$ and $\bi 3_d$ are $d$-dimensional vectors with each component being $0$, $1$ or $3$, respectively, and $\bi I_{d\times d}$ is the $d$-dimensional identity matrix. Here, we consider $d=5$. The multivariate exponential distribution is the multidimensional extension of the univariate exponential distribution.
The average coverage probabilities and average lengths of the coverage probabilities as well as their standard deviations (in parenthesis) for 95\% confidence intervals are provided in Table \ref{tab:Table3}. 
\begin{table}[H]
\caption{Coverage probabilities (standard deviations) and average lengths (standard deviations) of the categorical Gini correlations' interval estimators under different distributions when $K=2$ and $d=5$.}
\center
\scriptsize
\renewcommand{\arraystretch}{1.1}
\begin{tabular}{lllcccccccccr }
\hline \hline
Distribution&Parameter&{Method}& \multicolumn{2}{c}{$n=100$}& \multicolumn{2}{c} {$n=200$} \\
&&&CovProb & Length &CovProb & Length\\
\hline

&&JEL &.9463(.0059) &.0736(.0003) &.9508(.0061)&.0516(.0002)\\

$\frac{1}{2}\mathcal{N}(\bi 0_5, \bi I_{5\times 5})+\frac{1}{2}\mathcal{N}(\bi 1_5, 2 \bi I_{5\times 5})$
& 0.0803 &AJEL &.9515(.0062) & .0468(.0003) &.9537(.0063)&.0315(.0001)\\

&  &WJEL &.9427(.0074) & .0764(.0002) &.9400(.0070)&.0523(.0002)\\

&&JV &.9472(.0059) & .0610(.0001) &.9502(.0066)&.0422(.0001)
\\ \cline{1-7}

&&JEL &.9586(.0056) &.0625(.0006) &.9570(.0073)&.0397(.0004)\\

$\dfrac{2}{5}\mathcal{N}(\bi 0_5, \bi I_{5\times 5})+\dfrac{3}{5}\mathcal{N}(\bi 1_5, 2 \bi I_{5\times 5})$
& 0.0750 &AJEL &.9626(.0051) & .0443(.0001) &.9595(.0074)&.0305(.0001)\\

&&WJEL &.9611(.0057) & .0753(.0003) &.9614(.0067)&.0521(.0002)\\

&&JV &.9587(.0059) & .0586(.0002) &.9574(.0072)&.0406(.0001)
\\ \cline{1-7}

&&JEL &.9691(.0057) &.0219(.0002) &.9659(.0069)&.0128(.0001)\\

$\frac{1}{2}\mathcal{N}(\bi 0_5, \bi I_{5\times 5})+\frac{1}{2}\mathcal{N}(\bi 0_5, 3\bi I_{5\times 5})$
& 0.0173 &AJEL &.9719(.0055) & .0176(.0001) &.9680(.0068)&.0105(.0000)\\

&&WJEL &.9759(.0048) & .0309(.0002) &.9614(.0051)&.0187(.0001)\\

&&JV &.9709(.0057) & .0233(.0001) &.9674(.0074)&.0139(.0001)
\\ \cline{1-7}

&&JEL &.9729(.0060) &.0193(.0002) &.9713(.0057)&.0104(.0001)\\

$\dfrac{2}{5}\mathcal{N}(\bi 0_5, \bi I_{5\times 5})+\dfrac{3}{5}\mathcal{N}(\bi 0_5, 3\bi I_{5\times 5})$
& 0.0158 &AJEL &.9758(.0055) & .0164(.0001) &.9737(.0056)&.0097(.0000)\\

&&WJEL &.9689(.0056) & .0295(.0002) &.9304(.0082)&.0175(.0001)\\

&&JV &.9740(.0057) & .0213(.0001) &.9725(.0055)&.0128(.0000)
\\ \cline{1-7}

&&JEL &.9608(.0047) &.0877(.0002) &.9599(.0075)&.0611(.0001)\\

$\dfrac{1}{2}\textrm{Exp}(\bi 1_5)+\dfrac{1}{2}\textrm{Exp}(\bi 3_5)$ 
& 0.1274 &AJEL &.9656(.0047) & .0781(.0008) &.9627(.0076)&.0522(.0003)\\

&&WJEL &.9729(.0052) & .1094(.0005) &.9724(.0060)&.0770(.0003)\\

&&JV &.9621(.0056) & .0706(.0002) &.9590(.0078)&.0490(.0001)
\\ \cline{1-7}

&&JEL &.9768(.0055) &.0781(.0005) &.9778(.0053)&.0494(.0004)\\

$\dfrac{2}{5}\textrm{Exp}(\bi 1_5)+\dfrac{3}{5}\textrm{Exp}(\bi 3_5)$ 
& 0.1130 &AJEL &.9799(.0043) & .0529(.0002) &.9791(.0050)&.0359(.0001)\\

&&WJEL &.9784(.0051) & .1093(.0005) &.9729(.0042)&.0762(.0003)\\

&&JV &.9775(.0050) & .0675(.0001) &.9777(.0049)&.0470(.0001)
\\ \cline{1-7}

\hline\hline
\end{tabular}
\label{tab:Table3}
\end{table}
From Table \ref{tab:Table3}, we observe that under the first scenario, all the methods perform well and AJEL performs best with shortest confidence intervals. For the other two cases, all methods have over-coverage problem and JEL is superior to JV with better coverage probabilities and shorter confidence intervals when data is from the mixture of normal distributions. For all methods, the confidence interval gets shorter as the sample size becomes larger.

At the end of the simulation study, we consider $K=3, d=5$. The following scenarios with balanced $\bi p = (p_1, p_2, p_3) = (1/3, 1/3, 1/3)$ and unbalanced $\bi p = (4/12, 3/12, 5/12)$ of the total sample sizes of ($n = 120, n = 240$) are considered. 
\begin{itemize}
\item $\bi X \sim p_1\mathcal{N}(\bi 0_{5}, \bi I_{5\times 5})+p_2\mathcal{N}(\bi 1_5,   \bi I_{5\times 5})+p_3\mathcal{N}(\bi 2_5, \bi I_{5\times 5})$,
\item $\bi X \sim p_1\mathcal{N}(\bi 0_{5}, \bi I_{5\times 5})+p_2\mathcal{N}(\bi 0_{5}, 2\bi I_{5\times 5})+p_3\mathcal{N}(\bi 0_{5}, 3\bi I_{5\times 5})$,
\item $\bi X \sim p_1\textrm{Exp}(\bi 1_5)+p_2\textrm{Exp}(\bi 2_5)+p_3\textrm{Exp}(\bi 4_5)$.
\end{itemize}

%

The average coverage probabilities and average lengths as well as their standard deviations (in parenthesis) of 95\% confidence intervals are presented in Table \ref{tab:Table4}.

\begin{table}[H]
\caption{Coverage probabilities (standard deviations) and average lengths (standard deviations) of the categorical Gini correlations' interval estimators under different distributions when $K=3$ and $d=5$.}
\center
\scriptsize

\begin{adjustbox}{width=1\textwidth}
\renewcommand{\arraystretch}{1.1}
\begin{tabular}{lllcccccccccr }
\hline \hline
Distribution&Parameter&{Method}& \multicolumn{2}{c}{$n=120$}& \multicolumn{2}{c} {$n=240$} \\
&&&CovProb & Length &CovProb & Length\\
\hline

&&JEL &.9747(.0051) &.0736(.0007) &.9766(.0041)&.0466(.0004)\\
$\dfrac{1}{3}\mathcal{N}(\bi 0_{5}, \bi I_{5\times 5})+\dfrac{1}{3}\mathcal{N}(\bi 1_{5}, \bi I_{5\times 5})+\dfrac{1}{3}\mathcal{N}(\bi 2_{5}, \bi I_{5\times 5})$
& 0.2146 &AJEL &.9774(.0046)& .0591(.0002) &.9780(.0043)&.0407(.0001)\\
&&WJEL &.9709(.0045) & .0938(.0004) &.9708(.0048)&.0658(.0002)\\
&&JV &.9737(.005) & .0777(.0001) &.9765(.0037)&.0541(.0001)
\\ \cline{1-7}

&&JEL &.9800(.0048) &.0689(.0005) &.9813(.0045)&.0436(.0002)\\
$\dfrac{1}{4}\mathcal{N}(\bi 0_{5}, \bi I_{5\times 5})+\dfrac{1}{3}\mathcal{N}(\bi 1_{5}, \bi I_{5\times 5})+\dfrac{5}{12}\mathcal{N}(\bi 2_{5}, \bi I_{5\times 5})$ 
& 0.2081 &AJEL &.9823(.0041)& .0607(.0002) &.9826(.0044)&.0419(.0001)\\
&&WJEL &.9747(.0058) & .0965(.0004) &.9715(.0056)&.0675(.0002)\\
&&JV &.9800(.0046) & .0796(.0001) &.9813(.0047)&.0554(.0001)
\\ \cline{1-7}

&&JEL &.9804(.0050) &.0197(.0002) &.9873(.0039)&.0103(.0000)\\
$\dfrac{1}{3}\mathcal{N}(\bi 0_{5}, \bi I_{5\times 5})+\dfrac{1}{3}\mathcal{N}(\bi 0_{5}, 2\bi I_{5\times 5})+\dfrac{1}{3}\mathcal{N}(\bi 0_{5}, 3\bi I_{5\times 5})$ 
&0.0117&AJEL &.9820(.0047)& .0188(.0001) &.9795(.0040)&.0103(.0000)\\
&&WJEL &.9899(.0029) & .0309(.0002) &.9771(.0046)&.0172(.0001)\\
&&JV &.9807(.0049) & .0246(.0001) &.9874(.0041)&.0136(.0001)
\\ \cline{1-7}

&&JEL &.9817(.0053) &.0188(.0002) &.9803(.0046)&.0097(.0000)\\
$\dfrac{1}{4}\mathcal{N}(\bi 0_{5}, \bi I_{5\times 5})+\dfrac{1}{3}\mathcal{N}(\bi 0_{5}, 2\bi I_{5\times 5})+\dfrac{5}{12}\mathcal{N}(\bi 0_{5}, 3\bi I_{5\times 5})$ 
& 0.0104 &AJEL &.9839(.0047)& .0177(.0001) &.9817(.0045)&.0097(.0000)\\
&&WJEL &.9888(.0038) & .0299(.0002) &.9685(.0052)&.0164(.0001)\\
&&JV &.9819(.0051) & .0231(.0001) &.9803(.0047)&.0127(.0001)
\\ \cline{1-7}

&&JEL &.9670(.0064) &.0894(.0003) &.9670(.0064)&.0624(.0001)\\
$\dfrac{1}{3}\textrm{Exp}(\bi 1_5)+\dfrac{1}{3}\textrm{Exp}(\bi 2_5)+\dfrac{1}{3}\textrm{Exp}(\bi 4_5)$ 
& 0.1342 &AJEL &.9705(.0060)& .0728(.0007) &.9690(.0067)&.0482(.0005)\\
&&WJEL &.9795(.0051) & .1053(.0003) &.9813(.0046)&.0728(.0002)\\
&&JV &.9668(.0065) & .0728(.0001) &.9656(.0064)&.0502(.0001)
\\ \cline{1-7}

&&JEL &.9739(.0055) &.0834(.0004) &.9727(.0054)&.0579(.0002)\\
$\dfrac{1}{4}\textrm{Exp}(\bi 1_5)+\dfrac{1}{3}\textrm{Exp}(\bi 2_5)+\dfrac{5}{12}\textrm{Exp}(\bi 4_5)$ 
& 0.1219 &AJEL &.9767(.0052)& .0607(.0005) &.9744(.0050)&.0378(.0003)\\
&&WJEL &.9854(.0040) & .1045(.0004) &.9845(.0050)&.0731(.0002)\\
&&JV &.9721(.0052) & .0687(.0002) &.9713(.0051)&.0475(.0001)
\\ \cline{1-7}

\hline\hline
\end{tabular}
\end{adjustbox}
\label{tab:Table4}
\end{table}
From Table \ref{tab:Table4}, we can see that there is no significant difference among the coverage probabilities for all methods. However, our AJEL always has the shortest confidence intervals.

%

\section{Real data analysis}\label{sec:realdata}

For the purpose of illustration, we apply the proposed JEL method as well as the corresponding adjusted and robust JELs to two real data sets.

The first data set is the gilgai survey data (\cite{Jiang2011}, \cite{Sang2020}). This data set consists of 365 samples, which were taken at depths 0-10, 30-40 and 80-90 cm below the surface. Three features, pH, electrical conductivity (ec) in mS/cm and chloride content (cc) in ppm, are measured on a 1:5 soil:water extract from each sample. We treat the depths as the categorical $Y$ which takes three values, and the numeric variable $\bi X=(X_1, X_2, X_3)^T$ where each component denotes the features pH, ec, and cc respectively.
We use pH00 (0-10 cm), pH30 (30-40 cm), pH80 (80-90 cm), e00 (0-10 cm), e30 (30-40 cm), e80 (80-90 cm), and c00 (0-10 cm), c30 (30-40 cm) and c80 (80-90 cm) to denote pH, ec and cc at different depths, respectively. 
Our goal is to build confidence intervals for the categorical Gini correlation between the categorical variable `depth' and the numerical variable of the three features. 

\begin{figure}[H]
\caption{Density curves  of pH,  electrical conductivity and chloride content at different depth levels.}
\centering
\label{densityfig}
\begin{tabular}{cc}
\includegraphics[width=3.7in,height=1.9in]{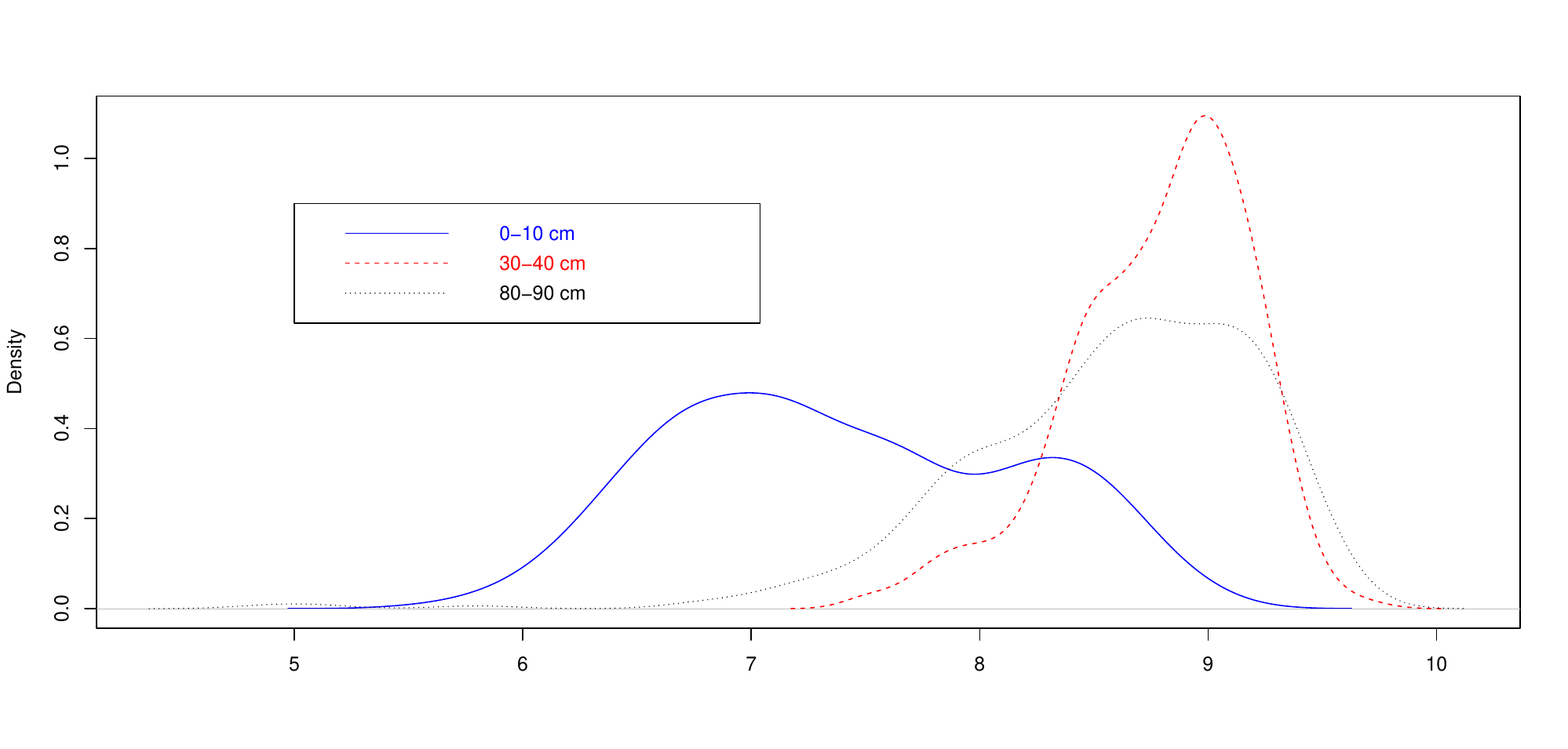} \vspace{-0.1in}\\
(a) density of pH\\ \vspace{0.1in}
\includegraphics[width=3.7in,height=1.9in]{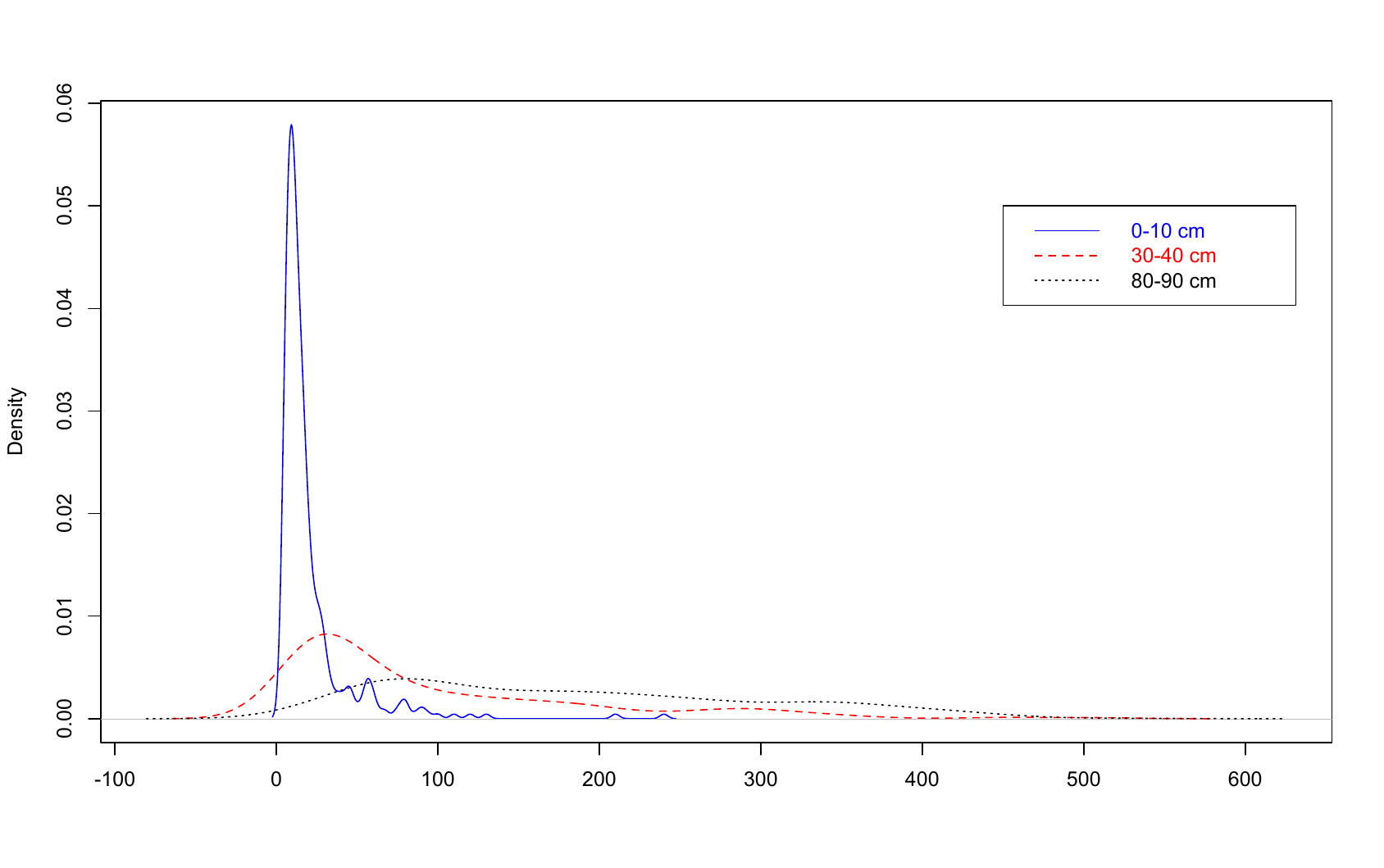} \vspace{-0.1in}\\
(b) density of ec\\ \vspace{0.1in}
\includegraphics[width=3.7in,height=1.9in ]{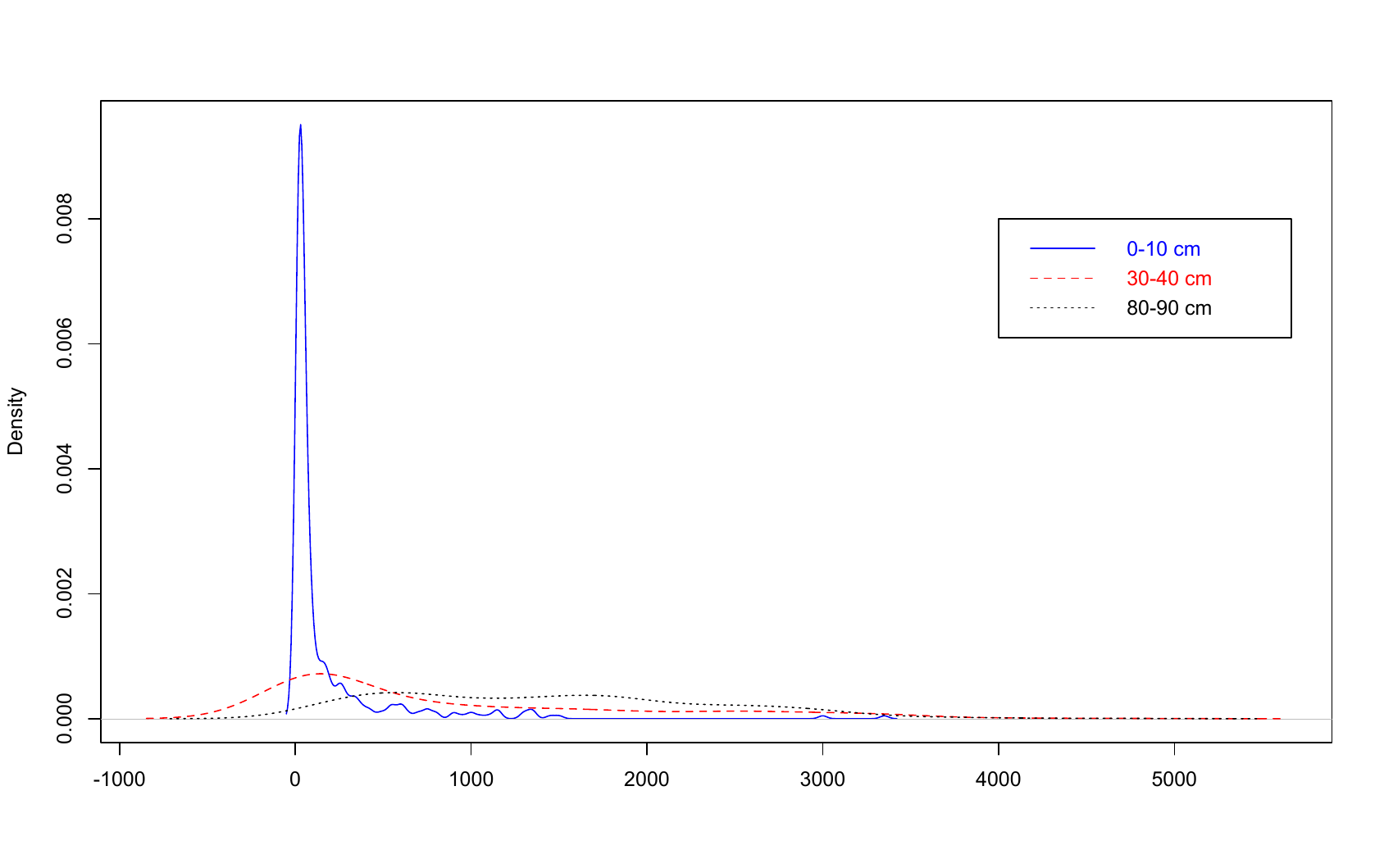} \vspace{-0.1in}\\
(c) density of cc \vspace{-0.1in} \\
\end{tabular}
\label{fig:density}
\end{figure}

We first visualize this data by drawing the density curves for each feature of $\bi X=(X_1, X_2, X_3)^T$ at different depth levels.
The density curves of pH,  ec and cc at different depth levels are drawn in Figure \ref{fig:density}. 

We observe that the distributions of each variable at different depths are quite different. The range and variation of each variable increase as the depth increases. At the same depth, ec and cc have a similar distribution although their scales are different. Those distributions are positively skewed, indicating the presence of a significant number of outliers in the two features at the 0-10 cm depth and only a few at the 30-40cm depth. 

The point estimates and confidence intervals for the categorical Gini correlations between depth levels and each of the three features, and also between depth levels and all features are listed in Table \ref{gil}. 
We compare the JEL, AJEL and WJEL with JV. The JV method is the inference method based on asymptotical normality with the asymptotical variance estimated by the jackknife method. 
\begin{table}[H]
\caption{Point estimates and 95\% confidence intervals for the categorical Gini correlations.}
\center
\label{gil}
\scriptsize
\begin{tabular}{lllcccccccccr }
\hline \hline
Method&Point estimate& Confidence interval \\

\hline
ph\\\hline

JEL & & (.2579, .3381)\\

AJEL &.3072&(.2579, .3382)\\

WJEL && (.2579, .3088)\\

JV &&(.2750, .3394)
\\ \hline
ee\\\hline

JEL & & (.2302, .3001)\\

AJEL &.2730&(.2587, .3002)\\

WJEL && (.2302, .2788)\\

JV &&(.2450, .3009)
\\ \hline

cc\\\hline

JEL & & (.1830, .2521)\\

AJEL &.2251&(.2111, .2522)\\

WJEL && (.1830, .2476)\\

JV &&(.1976, .2527)
\\ \hline

All\\\hline

JEL & & (.1827, .2514)\\

AJEL &.2247&(.2106, .2515)\\

WJEL && (.1827, .2510)\\

JV &&(.1973, .2520)
\\

\hline\hline
\end{tabular}
\label{tab:Table5}
\end{table}
The feature pH has the largest correlation with the depth levels, and the correlations for ee and cc with the depth levels are close. This finding is consistent with the plots in Figure \ref{densityfig} which shows larger difference among pH values at different depth values, and similar shapes of density curves for ec and cc. 
For the multivariate case, $d=3$, the AJEL performs best with  the shortest confidence interval.  For the univariate feature ee and cc, AJEL also generates the shortest confidence intervals and WJEL has the shortest one for pH.

The second data set is the famous Iris data set with the measurement in centimeters on Sepal Length, Sepal Width, Petal Length and Petal Width. The data set contains 3 species (Setosa,  Versicolour and Virginica) of 50 instances each.

\begin{figure}[H]
\centering
\label{scatter}
\begin{tabular}{cc}
\includegraphics[width=6.5in,height=5.5in]{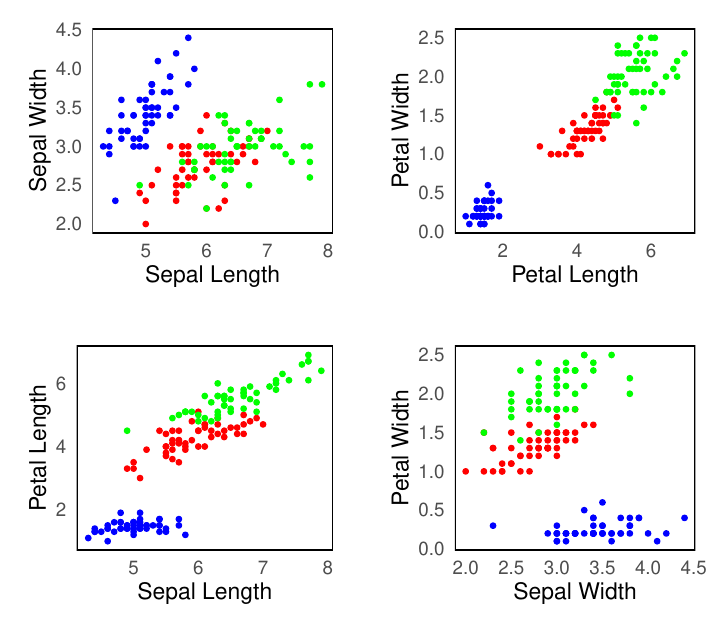} 
\end{tabular}
\caption{Scatter plots for Iris data. Blue, red and green are for Setosa, Versicolor, and Virginica, respectively.}
\label{fig:scatter}
\end{figure}

Figure \ref{fig:scatter} shows the scatter plots between selected pairs of two features and species. One can see that Versicolor and Virginica are close, and they can be easily distinguished from Setosa, except in the case of (Sepal Length, Sepal Width).

Table \ref{ci:iris} reports the point estimates and confidence intervals for the categorical Gini correlations between selected pairs of two features and species, as well as between all four features and species.

\begin{table}[H]
\caption{Point estimates and 95\% confidence intervals for categorical Gini correlations.}
\center
\scriptsize
\label{ci:iris}
\begin{tabular}{lllcccccccccr }
\hline \hline
Method&Point estimate& Confidence interval \\

\hline
(Sepal Length, Sepal Width)\\\hline

JEL & & (.3245, .4043)\\
AJEL &.3570&(.3245, .4051)\\
WJEL && (.2595, .3933)\\
JV &&(.3064, .4076)
\\ \hline

(Petal Length, Petal Width)\\\hline

JEL & & (.7344, .7878)\\
AJEL &.7561&(.7344, .7883)\\
WJEL && (.6911, .7720)\\
JV &&(.7223, .7899)
\\ \hline

(Sepal Length, Petal Length)\\\hline

JEL & & (.6337, .6972)\\
AJEL &.6596&(.6337, .6979)\\
WJEL && (.6337, .7068)\\
JV &&(.6193, .6999)
\\ \hline

(Sepal Width, Petal Width)\\\hline

JEL & & (.5311, .5954)\\
AJEL &.5572&(.5311, .5960)\\
WJEL && (.5311, .6040)\\
JV &&(.5165, .5980)
\\ \hline

All\\\hline

JEL & & (.6003, .6587)\\
AJEL &.6239&(.6003, .6593)\\
WJEL && (.6003, .6803)\\
JV &&(.5871, .6607)
\\ \hline

\hline\hline
\end{tabular}
\label{tab:Table5}
\end{table}

As expected, (Sepal Length, Sepal Width) has the smallest correlation and (Petal Length, Petal Width) has the largest correlation with species. 
JEL and AJEL perform closely with similar lengths of confidence intervals.
However, JEL has the shortest confidence interval in all cases.

\section{Conclusions and future work}\label{sec:conclusion}

In this paper, we define an estimating equation for the Categorical Gini correlation in the form of a function of U-statistics and develop the jackknife empirical likelihood. We further derive the adjusted jackknife empirical likelihood and Weighted jackknife empirical likelihood.

We establish the asymptotic properties of the jackknife empirical likelihood and
assess the empirical performance of the resulting interval estimators. Simulation studies suggest that our JEL interval estimators are competitive to existing methods in terms of coverage accuracy and shortness of confidence intervals. For multivariate cases, some cases result in conservative coverage probabilities. Future research could theoretically focus on improving the performance of those cases. 

Sang and Dang \cite{Sang2023} just developed a project to screen grouped features for ultrahigh-dimensional classification by using the categorical Gini correlation. The importance of the grouped/marginal feature is ranked by the order of the categorical Gini correlation with the categorical response. Thus, comparing the Gini correlations will be of interest. 
In the future, we will focus on extending the categorical Gini correlation for novel statistical applications.

\section{Appendix}
\noindent \textbf{Proof of Theorem \ref{wilkrho}}

The proof of Theorem \ref{wilkrho}  is closely related to the lemmas and corollaries from \cite{Jing2009}.  Hence, we provide a proof of Theorem \ref{wilkrho}
by checking the conditions  in \cite{Jing2009}.



First of all, we can show that 
\begin{align}
 W_n(\rho_g)=\dfrac{1}{n}\sum_{i=1}^n \hat{V}_i(\rho_g). \label{eqn:w}
 \end{align}

Without loss of generality, we assume $K=3$, and the data has been sorted in Y.
\begin{align*}
W_n(\rho_g)&=U_n(1-\rho_g)-\sum_{k=1}^3 \hat{p}_kU_{n_k}\\
&=(1-\rho_g) {n \choose 2}^{-1}\sum_{1\leq i<j \leq n}\|\bi X_i-\bi X_j\|-\sum_{k=1}^3 \hat{p}_k {n_k \choose 2}^{-1}\sum_{1\leq i<j \leq n_k}\|\bi X^{(k)}_i-\bi X^{(k)}_j\|.
\end{align*}
\begin{enumerate}
\item For $1 \leq i \leq n_1$,
\begin{align*}
W^{(-i)}_n(\rho_g)
&=(1-\rho_g) {n-1 \choose 2}^{-1}\sum_{1\leq j<l \leq n, j, l \neq i}\|\bi X_j-\bi X_l\|-\hat{p}_1 {n_1-1 \choose 2}^{-1}\sum_{1 \leq j <l\leq n_1, j, l \neq i}\|\bi X^{(1)}_j-\bi X^{(1)}_l\|\\
&- \hat{p}_2 {n_2 \choose 2}^{-1}\sum_{1\leq i<j \leq n_2}\|\bi X^{(2)}_i-\bi X^{(2)}_j\|-\hat{p}_3 {n_3 \choose 2}^{-1}\sum_{1\leq i<j \leq n_3}\|\bi X^{(3)}_i-\bi X^{(3)}_j\|.
\end{align*}
\item When $n_1+1 \leq i \leq n_1+n_2$,
\begin{align*}
W^{(-i)}_n(\rho_g)
&=(1-\rho_g) {n-1 \choose 2}^{-1}\sum_{1\leq j<l \leq n, j, l \neq i}\|\bi X_j-\bi X_l\|-\hat{p}_1 {n_1 \choose 2}^{-1}\sum_{1 \leq i<j \leq n_1}\|\bi X^{(1)}_i-\bi X^{(1)}_j\|\\
&-\hat{p}_2 {n_2-1 \choose 2}^{-1}\sum_{1 \leq j <l\leq n_2, j, l \neq i}\|\bi X^{(2)}_j-\bi X^{(2)}_l\|- \hat{p}_3 {n_3 \choose 2}^{-1}\sum_{1\leq i<j \leq n_3}\|\bi X^{(3)}_i-\bi X^{(3)}_j\|.
\end{align*}
\item For $n_1+n_2+1 \leq i \leq n$,
\begin{align*}
W^{(-i)}_n(\rho_g)
&=(1-\rho_g) {n-1 \choose 2}^{-1}\sum_{1\leq j<l \leq n, j, l \neq i}\|\bi X_j-\bi X_l\|-\hat{p}_1 {n_1 \choose 2}^{-1}\sum_{1 \leq i<j \leq n_1}\|\bi X^{(1)}_i-\bi X^{(1)}_j\|\\
&-\hat{p}_2 {n_2 \choose 2}^{-1}\sum_{1 \leq i <j \leq n_2}\|\bi X^{(2)}_j-\bi X^{(2)}_l\|- \hat{p}_3 {n_3-1 \choose 2}^{-1}\sum_{1\leq j<l \leq n_3, j, l \neq i}\|\bi X^{(3)}_i-\bi X^{(3)}_j\|.
\end{align*}
\end{enumerate}

Therefore,
\begin{enumerate}
\item for $1 \leq i \leq n_1$, we have 
\begin{align}
\hat{V}_i(\rho_g)&=nW_n(\rho_g)-(n-1)W^{(-i)}_n(\rho_g)\nonumber\\
&=(1-\rho_g)\left( n{n \choose 2}^{-1}\sum_{1\leq i<j \leq n}\|\bi X_i-\bi X_j\|-(n-1){n-1 \choose 2}^{-1}\sum_{1\leq j<l \leq n, j, l \neq i}\|\bi X_j-\bi X_l\| \right)\nonumber\\
&-\hat{p}_1\left( n{n_1 \choose 2}^{-1}\sum_{1\leq i<j \leq n_1}\|\bi X^{(1)}_i-\bi X^{(1)}_j\|-(n-1){n_1-1 \choose 2}^{-1}\sum_{1\leq j<l \leq n_1, j, l \neq i}\|\bi X^{(1)}_j-\bi X^{(1)}_l\| \right)\nonumber\\
&-\hat{p}_2 {n_2 \choose 2}^{-1}\sum_{1\leq i<j \leq n_2}\|\bi X^{(2)}_i-\bi X^{(2)}_j\|-\hat{p}_3 {n_3 \choose 2}^{-1}\sum_{1\leq i<j \leq n_3}\|\bi X^{(3)}_i-\bi X^{(3)}_j\|; \label{v1}
\end{align}
\item for $n_1+1 \leq i \leq n_1+n_2$, we have obtained
\begin{align}
\hat{V}_i(\rho_g)&=nW_n(\rho_g)-(n-1)W^{(-i)}_n(\rho_g)\nonumber\\
&=(1-\rho_g)\left( n{n \choose 2}^{-1}\sum_{1\leq i<j \leq n}\|\bi X_i-\bi X_j\|-(n-1){n-1 \choose 2}^{-1}\sum_{1\leq j<l \leq n, j, l \neq i}\|\bi X_j-\bi X_l\| \right)\nonumber\\
&-\hat{p}_2\left( n{n_2 \choose 2}^{-1}\sum_{1\leq i<j \leq n_2}\|\bi X^{(2)}_i-\bi X^{(2)}_j\|-(n-1){n_2-1 \choose 2}^{-1}\sum_{1\leq j<l \leq n_2, j, l \neq i}\|\bi X^{(2)}_j-\bi X^{(2)}_l\| \right)\nonumber\\
&-\hat{p}_1 {n_1 \choose 2}^{-1}\sum_{1\leq i<j \leq n_1}\|\bi X^{(1)}_i-\bi X^{(1)}_j\|-\hat{p}_3 {n_3 \choose 2}^{-1}\sum_{1\leq i<j \leq n_3}\|\bi X^{(3)}_i-\bi X^{(3)}_j\|; \label{v2}
\end{align}
\item when $n_1+n_2+1 \leq i \leq n$, we have 
\begin{align}
\hat{V}_i(\rho_g)&=nW_n(\rho_g)-(n-1)W^{(-i)}_n(\rho_g)\nonumber\\
&=(1-\rho_g)\left( n{n \choose 2}^{-1}\sum_{1\leq i<j \leq n}\|\bi X_i-\bi X_j\|-(n-1){n-1 \choose 2}^{-1}\sum_{1\leq j<l \leq n, j, l \neq i}\|\bi X_j-\bi X_l\| \right)\nonumber\\
&-\hat{p}_3\left( n{n_3 \choose 2}^{-1}\sum_{1\leq i<j \leq n_3}\|\bi X^{(3)}_i-\bi X^{(3)}_j\|-(n-1){n_3-1 \choose 2}^{-1}\sum_{1\leq j<l \leq n_3, j, l \neq i}\|\bi X^{(3)}_j-\bi X^{(3)}_l\| \right)\nonumber\\
&-\hat{p}_1 {n_1 \choose 2}^{-1}\sum_{1\leq i<j \leq n_1}\|\bi X^{(1)}_i-\bi X^{(1)}_j\|-\hat{p}_2 {n_2 \choose 2}^{-1}\sum_{1\leq i<j \leq n_2}\|\bi X^{(2)}_i-\bi X^{(2)}_j\|. \label{v3}
\end{align}
\end{enumerate}
By combining (\ref{v1}), (\ref{v2}) and ({\ref{v3}}), we have
\begin{align*}
\sum_{i=1}^n \hat{V}_i(\rho_g)&=(1-\rho_g)\sum_{i=1}^n \left (nU_n-(n-1)U^{(-i)}_{n-1}\right )\\
&-\hat{p}_1 \left\{\sum_{i=1}^{n_1}\left (n U_{n_1}-(n-1)U^{(-i)}_{n_1-1}\right )+(n_2+n_3)U_{n_1} \right \}\\
&-\hat{p}_2 \left\{\sum_{i=1}^{n_2}\left (n U_{n_2}-(n-1)U^{(-i)}_{n_2-1}\right )+(n_1+n_3)U_{n_2} \right \}\\
&-\hat{p}_3 \left\{\sum_{i=1}^{n_3}\left (n U_{n_3}-(n-1)U^{(-i)}_{n_3-1}\right )+(n_1+n_2)U_{n_3} \right \}\\
&=(1-\gamma) n U_n\\
&-\hat{p}_1 \left\{n_1n U_{n_1}-(n-1)\sum_{i=1}^{n_1}U^{(-i)}_{n_1-1}+(n_2+n_3)U_{n_1} \right \}\\
&-\hat{p}_2 \left\{n_2n U_{n_2}-(n-1)\sum_{i=1}^{n_2}U^{(-i)}_{n_2-1}+(n_1+n_3)U_{n_2} \right \}\\
&-\hat{p}_1 \left\{n_3n U_{n_3}-(n-1)\sum_{i=1}^{n_3}U^{(-i)}_{n_3-1}+(n_1+n_2)U_{n_3} \right \}\\
&=(1-\rho_g) n U_n-\hat{p}_1 n U_{n_1}-\hat{p}_2 n U_{n_2}-\hat{p}_3 n U_{n_3}\\
&=nW_n(\rho_g)
\end{align*}
by the fact that 
\begin{align*}
nU_n=\sum_{i=1}^n U^{(-i)}_{n-1}, \ n_1U_{n_1}=\sum_{i=1}^{n_1} U^{(-i)}_{n_1-1}.
\end{align*}
Thus, we have $W_n(\rho_g)=\dfrac{1}{n}\sum_{i=1}^n \hat{V}_i(\rho_g)$.


Next, we will show that $W_n(\rho_g)$ is not degenerate and hence has a normal limit under condition \textbf{C}2. 
The calculation of $W_n(\rho_g)$ involves complicated constants. As such, we will show the normal limit for $W_n(\rho_g)$ by providing the normal limit of the corresponding $V$-statistic,
\begin{align}\label{V-est}
G_n(\rho_g)&=\dfrac{1}{n^2} \sum_{i=1}^n \sum_{j=1}^n \|\bi X_i-\bi X_j\|(1-\rho_g)-\sum_{k=1}^K \hat{p}_k\dfrac{1}{n^2_k} \sum_{i_1=1}^{n_k}\sum_{j_1=1}^{n_k}\|\bi X^{(k)}_i-\bi X^{(k)}_j\|.
\end{align}

It has been shown that (\cite{Dang2021})
\begin{align*}
&\Delta=\sum_{k=1}^K p^2_k \Delta_k+2\sum_{1 \leq k<l \leq K}p_kp_l \Delta_{kl},\\
&\textrm{gCov}(\bi X, Y)=\sum_{k=1}^K T(\bi X^{(k)}, \bi X)=\sum_{1 \leq k<l \leq K}p_kp_lT(\bi X^{(k)}, \bi X^{(l)}),\\
\end{align*}
where $T(\bi X^{(k)}, \bi X)=2\E\|\bi X^{(k)}-\bi X\|-\E\|\bi X^{(k)}-\bi X^{(k)'}\|-\E\|\bi X-\bi X'\|$.
Then we have
\begin{align*}
\textrm{gCov}(\bi X, Y)-\rho_g \Delta&=\sum_{ 1\leq k<l \leq K}p_kp_l\{2 \Delta_{kl}-\Delta_k-\Delta_l\}-\rho_g [\sum_{k=1}^K p^2_k \Delta_k+2\sum_{1 \leq k<l \leq K}p_kp_l \Delta_{kl}]\\
&=\sum_{1 \leq k<l \leq K} p_kp_l  \left\{2(1-\rho_g)\Delta_{kl} -(1+\rho_g \dfrac{p_k}{1-p_k})\Delta_k-(1+\rho_g \dfrac{p_l}{1-p_l}\Delta_l\right\},
\end{align*}
where the last equality holds because 
\begin{align*}
\sum_{k=1}^K p^2_k \Delta_k=\sum_{1 \leq k<l\leq K}p_kp_l(\dfrac{p_k}{1-p_k}\Delta_k+\dfrac{p_l}{1-p_l}\Delta_l).
\end{align*}

Let
\begin{align*}
\Gamma_n(\rho_g)&=\sum_{1 \leq k < l \leq K}\hat{p_k}\hat{p_l}\hat{\Gamma}_{kl}(\rho_g),
\end{align*}
where
\begin{align*}
\hat{\Gamma}_{kl}(\rho_g)
&= 2 (1-\rho_g) \dfrac{1}{n_k n_l}\sum_{i=1}^{n_k}\sum_{j=1}^{n_l}\|\bi X^{(k)}_i-\bi X^{(l)}_j\|\\
&-\big(1+\dfrac{\rho_g\hat{p_k}}{1-\hat{p}_k }\big)\dfrac{1}{n^2_k}\sum_{i_1=1}^{n_k}\sum_{i_2=1}^{n_k}\|\bi X^{(k)}_{i_1}-\bi X^{(k)}_{i_2}\|\\
&-\big(1+\dfrac{\rho_g\hat{p_l}}{1-\hat{p}_l}\big)\dfrac{1}{n^2_l}\sum_{j_1=1}^{n_l}\sum_{j_2=1}^{n_l}\|\bi X^{(l)}_{j_1}-\bi X^{(l)}_{j_2}\|\\
&:=\dfrac{1}{n^2_k} \dfrac{1}{n^2_l}\sum_{i_1, i_2=1}^{n_k}\sum_{j_1, j_2=1}^{n_l}g(\bi X^{(k)}_{i_1}, \bi X^{(k)}_{i_2}; \bi X^{(l)}_{j_1}, \bi X^{(l)}_{j_2})
\end{align*}
with
\begin{align*}
g_{kl}:=g(\bi X^{(k)}_{i_1}, \bi X^{(k)}_{i_2}; \bi X^{(l)}_{j_1}, \bi X^{(l)}_{j_2})&= (1-\rho_g)\|\bi X^{(k)}_{i_1}-\bi X^{(l)}_{j_1}\|+ (1-\rho_g)\|\bi X^{(k)}_{i_2}-\bi X^{(l)}_{j_2}\|\\
&-\big(1+\dfrac{\rho_g\hat{p_k}}{1-\hat{p}_k}\big)\|\bi X^{(k)}_{i_1}-\bi X^{(k)}_{i_2}\|-\big(1+\dfrac{\rho_g\hat{p_l}}{1-\hat{p}_l }\big)\|\bi X^{(l)}_{j_1}-\bi X^{(l)}_{j_2}\|.
\end{align*}
In fact, the estimator $\Gamma_n(\rho_g)$ is the same as the V-statistic estimator $G_n(\rho_g)$ in (\ref{V-est}) becasue
\begin{align*}
\Gamma_n(\rho_g)&=\sum_{1 \leq k < l \leq K}\hat{p_k}\hat{p_l}\hat{\Gamma}_{kl}(\rho_g)\\
&=\dfrac{1}{n^2}\sum_{ 1\leq k\neq l \leq K} (1-\rho_g) \sum_{i=1}^{n_k}\sum_{j=1}^{n_l}\|\bi X^{(k)}_{i}-\bi X^{(l)}_{j}\|-\sum_{k=1}^K \hat{p}_k(1-\hat{p}_k) \dfrac{1}{n^2_k}\sum_{i_1=1}^{n_k}\sum_{i_2=1}^{n_k}\|\bi X^{(k)}_{i_1}-\bi X^{(k)}_{i_2}\|\\
&-\rho_g \sum_{k=1}^K \hat{p}^2_k \dfrac{1}{n^2_k}\sum_{i_1=1}^{n_k}\sum_{i_2=1}^{n_k}\|\bi X^{(k)}_{i_1}-\bi X^{(k)}_{i_2}\|\\
&=\dfrac{1}{n^2}\sum_{ 1\leq k\neq l \leq K} (1-\rho_g) \sum_{i=1}^{n_k}\sum_{j=1}^{n_l}\|\bi X^{(k)}_{i}-\bi X^{(l)}_{j}\|-\sum_{k=1}^K \hat{p}_k\dfrac{1}{n^2_k}\sum_{i_1=1}^{n_k}\sum_{i_2=1}^{n_k}\|\bi X^{(k)}_{i_1}-\bi X^{(k)}_{i_2}\|\\
&+(1-\rho_g) \sum_{k=1}^K \hat{p}^2_k \dfrac{1}{n^2_k}\sum_{i_1=1}^{n_k}\sum_{i_2=1}^{n_k}\|\bi X^{(k)}_{i_1}-\bi X^{(k)}_{i_2}\|\\
&=\dfrac{(1-\rho_g)}{n^2}\left\{\sum_{ 1\leq k\neq l \leq K}  \sum_{i=1}^{n_k}\sum_{j=1}^{n_l}\|\bi X^{(k)}_{i}-\bi X^{(l)}_{j}\|+\sum_{k=1}^K \sum_{i_1=1}^{n_k}\sum_{i_2=1}^{n_k}\|\bi X^{(k)}_{i_1}-\bi X^{(k)}_{i_2}\|\right \}\\
&-\sum_{k=1}^K \hat{p}_k\dfrac{1}{n^2_k}\sum_{i_1=1}^{n_k}\sum_{i_2=1}^{n_k}\|\bi X^{(k)}_{i_1}-\bi X^{(k)}_{i_2}\|\\
&=G_n(\rho_g).
\end{align*}

Consider the centered kernel function $\tilde{g}_{kl}=g_{kl}-\mathbb{E}g_{kl}$ and its first-order projections as follows.
\begin{align*}
&\tilde{g}^{10}_{kl}(\bi x^{(k)})=\E \tilde{g}_{kl}(\bi x^{(k)}, \bi X^{(k)}_{2}; \bi X^{(l)}_{1}, \bi X^{(l)}_{2})\\
&=(1-\rho_g)\mathbb{E}\|\bi x^{(k)}-\bi X^{(l)}_1\|-(1+\rho_g \dfrac{p_k}{1-p_k})\mathbb{E}\|\bi x^{(k)}-\bi X^{(k)}_2\|-(1-\rho_g)\Delta_{kl}+(1+\rho_g \dfrac{p_k}{1-p_k})\Delta_k,\\
&\tilde{g}^{01}_{kl}(\bi x^{(l)})=\E \tilde{g}_{kl}(\bi X^{(k)}_1, \bi X^{(k)}_{2}; \bi x^{(l)}, \bi X^{(l)}_{2})\\
&=(1-\rho_g)\mathbb{E}\|\bi X^{(k)}_1-\bi x^{(l)}\|-(1+\rho_g \dfrac{p_l}{1-p_l})\mathbb{E}\|\bi x^{(l)}-\bi X^{(l)}_2\|-(1-\rho_g)\Delta_{kl}+(1+\rho_g \dfrac{p_l}{1-p_l})\Delta_l.
\end{align*}
Denote the first-order projection of $\hat{\Gamma}_{kl}(\rho_g)$ as $\hat{\Gamma}^{(1)}_{kl}(\rho_g)=2[\dfrac{1}{n_k}\sum_{i=1}^{n_k}\tilde{g}^{10}_{kl}(\bi X^{(k)}_{i})+\dfrac{1}{n_l}\sum_{j=1}^{n_l}\tilde{g}^{01}_{kl}(\bi X^{(l)}_{j})]$. 
Then 
\begin{align*}
\sum_{1 \leq k<l \leq K}\hat{\Gamma}^{(1)}_{kl}(\rho_g)=\dfrac{4}{n^2}\sum_{k \neq l}n_l\sum_{i=1}^{n_k}\tilde{g}^{10}_{kl}(\bi X^{(k)}_i).
\end{align*}
We can show that $\rho_g \neq 0$ if and only if at least one $\textrm{var}\big(\tilde{g}^{10}_{kl}(\bi X^{(k)}_i)\big)$ is not zero.  If at least one $\textrm{var}\big(\tilde{g}^{10}_{kl}(\bi X^{(k)}_i)\big)$ is not zero, then $\Delta_k=\Delta_l=\Delta_{kl}$ does not hold for all $k, l$, therefore, $\rho_g \neq 0$. 
On the other hand, under $\rho_g \neq 0$, if all $\textrm{var}\big(\tilde{g}^{10}_{kl}(\bi X^{(k)})\big)=0$, then $\tilde{g}^{10}_{kl}(\bi X^{(k)})$ is a constant which is $\E \tilde{g}^{10}_{kl}(\bi X^{(k)})$. It is easy to check $\E \tilde{g}^{10}_{kl}(\bi X^{(k)})=0$.  We also have  $\tilde{g}^{01}_{kl}(\bi X^{(l)})=0$. 
Consequently, 
\begin{align*}
&\tilde{g}^{10}_{kl}(\bi x^{(k)})=\left\{\mathbb{E}\|\bi x^{(k)}-\bi X^{(l)}_1\|-\mathbb{E}\|\bi x^{(k)}-\bi X^{(k)}_2\|-\Delta_{kl}+\Delta_k\right \}\\
&-\rho_g \left\{\mathbb{E}\|\bi x^{(k)}-\bi X^{(l)}_1\|+\dfrac{p_k}{1-p_k}\mathbb{E}\|\bi x^{(k)}-\bi X^{(k)}_2\|-\Delta_{kl}-  \dfrac{p_k}{1-p_k}\Delta_k\right \} \\
&=0,\\
&\tilde{g}^{01}_{kl}(\bi x^{(l)})=\left\{\mathbb{E}\|\bi X^{(k)}_1-\bi x^{(l)}\|-\mathbb{E}\|\bi x^{(l)}-\bi X^{(l)}_2\|-\Delta_{kl}+\Delta_l\right \}\\
&-\rho_g \left\{\mathbb{E}\|\bi X^{(k)}_{1}-\bi x^{(l)}\|+\dfrac{p_l}{1-p_l}\mathbb{E}\|\bi x^{(l)}-\bi X^{(l)}_2\|-\Delta_{kl}-  \dfrac{p_l}{1-p_l}\Delta_l\right \} \\
&=0,
\end{align*}
for all $k,l$.
This implies $\Delta_k=\Delta_l=\Delta_{kl}$ for all $k,l$, and hence $\rho_g=0$ which contradicts the fact that $\rho_g \neq 0$.  Thus, there exists at least one nonzero $\textrm{var}\big(\tilde{g}^{10}_{kl}(\bi X^{(k)}_i)\big)$. Hence, $\Gamma_n(\rho_g)$ converges to a normal distribution.
The $U$-estimator $W_n(\rho_g)$  and the $V$-estimator $\Gamma_n(\rho_g)$ are asymptotically equivalent. Therefore, $W_n(\rho_g)$ admits a normal limit.

Overall, $W_n(\rho_g)$ in (\ref{wn}) satisfies all the lemmas  for the JEL procedure in \cite{Jing2009}. 
This completes the proof.

\end{document}